# Estimation of the shape parameter of a generalized Pareto distribution based on a transformation to Pareto distributed variables

J. Martin van Zyl

**Abstract**  Random variables of the generalized three-parameter Pareto distribution, can be transformed to that of the Pareto distribution. Explicit expressions exist for the maximum likelihood estimators of the parameters of the Pareto distribution. The performance of the estimation of the shape parameter of generalized Pareto distributed using transformed observations is tested. It was found to improve the performance with respect to relative efficiency. This was also tested in some peak over threshold problems and good results were found.

## 1  Introduction

A transformation from a generalized three-parameter Pareto distribution (GPD) to a Pareto distribution (Pareto Type I distribution) with the same support and shape parameter is shown to exist. The three-parameter GPD is also called a Pareto type II distribution. The performance of the estimation of the tail index of the GPD using maximum likelihood estimation of the index on the Pareto transformed observations, will be investigated in this work. No explicit expressions exist for the maximum likelihood (ML) estimators of the GPD and numerical methods must be used to find the ML estimators and besides this there are some problems when using ML estimation for the GPD which was investigated by del Castillo and Daoudi (2009), Grimshaw (1993). It can happen in small samples that the ML estimators do not exist. Transformation using estimated parameters to improve and stabilize estimation is used in many problems in statistics. An example is where observations are standardized using initial estimated parameters before applying empirical characteristic regression models results in more stable estimation when estimating the parameters of a stable distribution (Paulson et al., 1975), Koutrouvellis (1980).

The density function of the three-parameter GPD is

$$f(x) = (1/\sigma)(1+(\xi/\sigma)(x-\mu))^{-1-1/\xi}, \; x \geq \mu, \sigma > 0, \xi > 0, \qquad (1)$$

$\sigma$ is a scale parameter, $\xi$ the shape parameter and $\mu$ the location parameter. The tail index is $\alpha = 1/\xi$ and the case where $0 < \xi \leq 1$ will be considered in this work. This is the range of $\alpha$ which is often of interest in financial applications, thus with possibly an infinite second moment.

Consider the transformation $z/\mu = 1 + (\xi/\sigma)(x-\mu)$ or

$$z = (\xi\mu/\sigma)x + \mu(1-\xi\mu/\sigma). \qquad (2)$$

Thus $x = (\sigma/\xi\mu)z + \mu - \sigma/\xi$ and the Jacobian for $\xi > 0$ of the transformation is $\sigma/\mu\xi$ and it follows that the density of the transformed variable is Pareto distributed with density

$$f(z) = \alpha(\mu^\alpha / z^{\alpha+1}), \; z \geq \mu, \alpha > 0. \qquad (3)$$

By letting $x = \mu$ in (2) it can be seen that the support of the Pareto is similar to that of the GPD. The tail index is the same as that of the original GPD. For a Pareto distributed observations as in (3), with an observed sample, $z_1, ..., z_n$, the maximum likelihood estimator of $\xi = 1/\alpha$ and $\mu$ are

$$\hat{\xi} = \frac{1}{n}\sum_{j=1}^{n}\log(z_j / \hat{\mu}), \; \hat{\mu} = z_{(1)}.$$

Suppose a sample of size n observations, $x_1, ..., x_n$, are available which are GPD distributed with parameters $\mu, \sigma, \xi$, the corresponding order statistics are $x_{(1)} \leq ... \leq x_{(n)}$. Initial estimate $\hat{\xi}, \hat{\sigma}$ and $\hat{\mu} = x_{(1)}$. Then in terms of the original sample:

$$\hat{\xi}_T = \frac{1}{n}\sum_{j=1}^{n}\log[(\hat{\xi}/\hat{\sigma})x_j + 1 - \hat{\xi}\hat{\mu}/\hat{\sigma}], \qquad (4)$$

and denote the estimated tail-index by $\hat{\alpha}_T = 1/\hat{\xi}_T$.

In practice the transformation involves unknown parameters and initial estimators of the parameters will be used to calculate the transformed observations and Pareto ML estimation can be applied to the transformed observations to estimate $\alpha$.

An alternative transformation when working with excesses over a threshold, or a two-parameter GPD with $\mu = 0$

$$f(x) = (1/\sigma)(1 + (\xi/\sigma)x)^{-1-1/\xi}, x \geq 0, \sigma > 0, \xi > 0,$$

is to let $z/\sigma = 1 + (\xi/\sigma)x$

$$f(z) = \alpha\sigma^{\alpha}/z^{\alpha+1}, \quad z \geq \sigma.$$

Related and similar properties are of the GPD is $(1/\xi)\log(1+\xi x/\sigma)$ has a standard exponential distribution for $\mu = 0$ and also $\log(z/\mu)$, for $z$ distributed as in (3), has an exponential distribution. A very specific case of the three-parameter distribution reduces to the Pareto distribution when the parameters are such that $\sigma = \xi\mu$ (McCulloch, 1997). This can also be seen by noting that $\bar{F}(x) = 1 - F(x)$, where F is the distribution function of the GPD can be written as

$$\begin{aligned}\bar{F}(x) &= (1+(\xi/\sigma)(x-\mu))^{-\xi}\\ &= x^{-1/\xi}((1-\mu\xi/\sigma)/x + \xi/\sigma)^{-1/\xi}\\ &= (\mu^{1/\xi}/x^{1/\xi})(\mu(1-\mu\xi/\sigma)/x + \xi\mu/\sigma)^{-1/\xi} \qquad (5)\end{aligned}$$

which implies that $\bar{F}(x) = (x/\mu)^{-1/\xi}, x \geq \mu$, for $\sigma = \mu\xi$. It can be seen that as $x \to \infty$, the GPD converges to a Pareto type I distribution too, even if $\mu\xi \neq \sigma$.

Percentiles and expected values of the original GPD can be found by calculating these values using the Pareto and then back transforming to GPD values, using the estimated parameters. For example consider the p-th percentile of the original GPD:

$$p = \int_{\mu}^{x_p} (1/\sigma)(1 + (\xi/\sigma)(x-\mu))^{-1-1/\xi} dx$$

Using the transformation (2) it follows that:

$$p = 1 - \left(\hat{\mu}/\hat{z}(x_p)\right)^{\hat{\alpha}_T},$$

$$z(x_p) = (\hat{\xi}\hat{\mu}/\hat{\sigma})x_p + \hat{\mu} - \hat{\xi}\hat{\mu}^2/\hat{\sigma}.$$

The success of the transformed estimator will depend on how well the initial estimation is. A review of estimation procedures for the GPD is given by de Zea, Bermudez, Kotz (2010). Two approaches will be used, one the simple and easy to calculate probability weighted method (PWM) and the other the Zhang and Stephens (2009) procedure based on empirical Bayes methods.

De Zea, Bermudez, Kotz (2010) found that the probability weighted method (PWM) method performs well for $0 \leq \xi \leq 1$ and very good for $\xi \leq 0.5$. Castillo and Hadi (1997) also showed that the estimator performs good in small samples for $\xi \leq 0.5$. Zhang and Stephens (2009) derived an estimator and showed that it performs excellent with respect to bias and mean square error. These two estimation procedures will be used to perform the transformation and check the performance of the Pareto ML estimator. It was found that the transformation improves the relative efficiency in both cases. For $\xi \leq 0.5$ the PWM method can be used, but for $\xi > 0.5$ it would be better to use the Zhang and Stephens (2009) to calculate the transformed observations.

Both these procedures were derived to estimate the GPD with two parameters and assuming that $\mu = 0$. To apply it on samples from the three-parameter GPD with $\mu \geq 0$, the estimation was conducted on the sample $z_j = x_j - \hat{\mu},\ j = 1,...,n,\ \hat{\mu} = x_{(1)}$. Samples will be simulated from a three-parameter GPD with $\mu > 0$, and the estimation conducted on the excesses over the threshold which is the smallest observation in the sample.

## 2  Simulation study to test the performance on GPD samples

Samples were generated from a GPD, the smallest observation chosen as an estimator of the lower bound of the support, thus $\hat{\mu} = x_{(1)}$. Then using the excesses over the point $\hat{\mu}$, the transformed observations $z_{(j)} = x_{(j)} - \hat{\mu},\ j = 1,...,n$, were formed. Using these excesses as the sample, the following four estimators were compared:

The Zhang and Stephens (2009) empirical Bayes estimator, Pareto ML estimation on transformed observations using the estimators of Zhang and Stephens (2009) to perform the transformation to Pareto distributed variables. PWM estimation, Pareto ML estimation on transformed observations using the PWM estimators to perform the transformation to Pareto distributed variables. The Pareto estimator is biased and corrections can also be made for this in specific problems. In table 3 the performance of the estimator was checked when the data was actually Pareto and it was assumed that it is from a GPD.

The number of simulations in each case is m=1000, and MSE was calculated as $(1/m)\sum_{j=1}^{m}(\hat{\xi}_j - \xi)^2$ for the various estimators. The bias is calculated as the true parameter minus the average estimated parameter. The ML estimate for Pareto samples of the index, performs very weak when the sample is GPD distributed and these results were not included in table 1 and table 2.

The relative efficiency of $\hat{\xi}$ was estimated by making use of the results of Smith (1984) and in a similar was as the application of this result in Zhang and Stephens (2009). The asymptotic variance of the MLE of the GPD is given by $nVar(\hat{\xi}_{MLE}) = (1+\xi)^2$. The true parameter was used to calculate this variance and the ratio this variance to the estimated MSE was used to estimate relative efficiency.

The plotting point chosen to estimate the empirical distribution when using the PWM method was $\hat{F}(x_{(j)}) = (j - 0.35)/n, \ j=1,...,n$, for a sample of size n, as suggested by Landwehr et al. (1979). Probability weighted moment (PWM) estimators were tested in this study to estimate $\theta$ for given $\mu$ and this procedure yielded good results. Thus using the excesses over the point $\hat{\mu} = x_{(1)}$, the transformed observations, $z_{(j)} = x_{(j)} - \hat{\mu}, \ j=1,...,n$. Three sample sizes, $n=50, 100, 250,$ and three sets of parameters were considered, $\mu = 1, \sigma = 1, 2$ and a range of values for $\xi$. In the third case $\sigma$ was chosen each time such that $\sigma = \mu\xi$ with $\mu = 1$.

| | $\sigma = 1, \mu = 1$ $\xi$ | Zhang&Stephens MSE and Bias | | Transformed (Zhang&Stephens) MSE and Bias | | PWM MSE and Bias | | Transformed (PWM) MSE and Bias | |
|---|---|---|---|---|---|---|---|---|---|
| n=50 | 0.1 | 0.0274 | -0.0178 | 0.0182 | -0.0408 | 0.0290 | 0.0377 | 0.0118 | -0.0038 |
| | 0.25 | 0.0324 | -0.0118 | 0.0286 | -0.0154 | 0.0304 | 0.0484 | 0.0237 | 0.0371 |
| | 0.5 | 0.0451 | -0.0009 | 0.0436 | 0.0025 | 0.0394 | 0.0844 | 0.0384 | 0.0815 |
| | 0.75 | 0.0587 | 0.0228 | 0.0575 | 0.0299 | 0.0612 | 0.1745 | 0.0592 | 0.1730 |
| | 1.0 | 0.0754 | 0.0300 | 0.0739 | 0.0399 | 0.1131 | 0.2973 | 0.1134 | 0.3043 |
| n=100 | 0.1 | 0.0126 | -0.0092 | 0.0094 | -0.0195 | 0.0134 | 0.0188 | 0.0078 | 0.0013 |
| | 0.25 | 0.0164 | -0.0125 | 0.0156 | -0.0131 | 0.0151 | 0.0180 | 0.0140 | 0.0156 |
| | 0.5 | 0.0227 | 0.0080 | 0.0225 | 0.0096 | 0.0225 | 0.0578 | 0.0221 | 0.0558 |
| | 0.75 | 0.0293 | 0.0124 | 0.0290 | 0.0154 | 0.0334 | 0.1266 | 0.0314 | 0.1215 |
| | 1.0 | 0.0395 | 0.0258 | 0.0393 | 0.0307 | 0.0737 | 0.2436 | 0.0721 | 0.2463 |
| n=250 | 0.1 | 0.0049 | -0.0034 | 0.0044 | -0.0054 | 0.0052 | 0.0082 | 0.0043 | 0.0045 |
| | 0.25 | 0.0066 | -0.0061 | 0.0066 | -0.0059 | 0.0061 | 0.0066 | 0.0061 | 0.0064 |
| | 0.5 | 0.0089 | -0.0002 | 0.0089 | 0.0004 | 0.0096 | 0.0278 | 0.0093 | 0.0261 |
| | 0.75 | 0.0115 | 0.0017 | 0.0115 | 0.0029 | 0.0171 | 0.0794 | 0.0155 | 0.0750 |
| | 1.0 | 0.0158 | 0.0087 | 0.0157 | 0.0106 | 0.0466 | 0.1964 | 0.0448 | 0.1955 |

**Table 1** Results of estimators of $\xi$ of a GPD with $\sigma = 1, \mu = 1$. MSE and estimated bias based on 1000 samples. The results of using two methods to calculate the transformed observations on which Pareto estimation is performed is shown. The procedures are the Zhang and Stephens and PWM methods.

| | $\sigma=1, \mu=1$ $\xi$ | Zhang&Stephens Relative Efficiency | Transformed (Zhang&Stephens) Relative Efficiency | PWM Relative Efficiency | Transformed (PWM) Relative Efficiency |
|---|---|---|---|---|---|
| **n=50** | **0.1** | 0.8834 | 1.3327 | 0.8353 | 2.0535 |
| | **0.25** | 0.9656 | 1.0909 | 1.0275 | 1.3211 |
| | **0.5** | 0.9988 | 1.0325 | 1.1435 | 1.1710 |
| | **0.75** | 1.0440 | 1.0649 | 1.0003 | 1.0354 |
| | **1.0** | 1.0610 | 1.0830 | 0.7070 | 0.7054 |
| **n=100** | **0.1** | 0.9583 | 1.2877 | 0.9003 | 1.5598 |
| | **0.25** | 0.9507 | 1.0032 | 1.0369 | 1.1159 |
| | **0.5** | 0.9907 | 1.0001 | 1.0004 | 1.0183 |
| | **0.75** | 1.0465 | 1.0560 | 0.9182 | 0.9751 |
| | **1.0** | 1.0115 | 1.0180 | 0.5428 | 0.5545 |
| **n=250** | **0.1** | 0.9915 | 1.1024 | 0.9246 | 1.1129 |
| | **0.25** | 0.9426 | 0.9458 | 1.0180 | 1.0170 |
| | **0.5** | 1.0098 | 1.0140 | 0.9359 | 0.9709 |
| | **0.75** | 1.0614 | 1.0684 | 0.7172 | 0.7881 |
| | **1.0** | 1.0152 | 1.0200 | 0.3436 | 0.3574 |

**Table 2** Relative efficiency of estimators of $\xi$ of a GPD with $\sigma=1, \mu=1$ based on 1000 samples. The results of using two methods to calculate the transformed observations on which Pareto estimation is performed is shown.

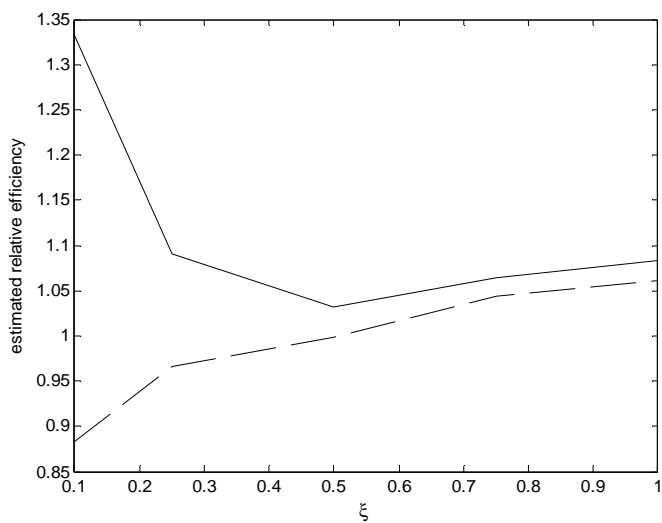

**Fig. 1** Plot of relative efficiency of estimators of $\xi$ of a GPD with $\sigma = 1, \mu = 1$ and n=50, based on 1000 samples. The dashed line is the relative efficiency using Zhang and Stephens (2009) and the solid line using these estimators to transform the observations and perform Pareto ML estimation.

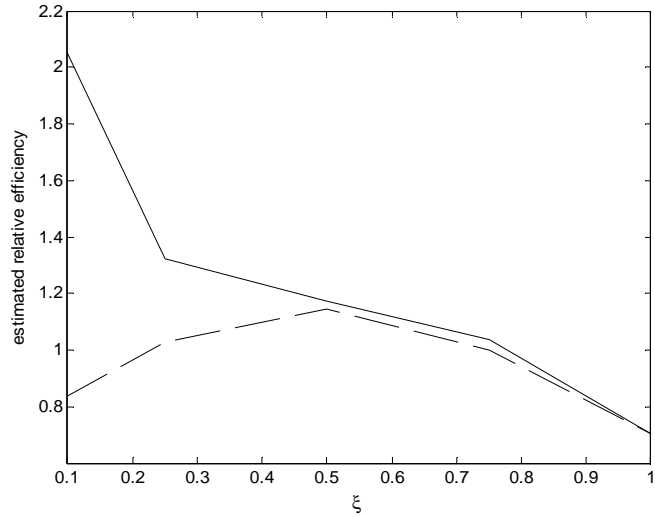

**Fig. 2** Plot of relative efficiency of estimators of $\xi$ of a GPD with $\sigma = 1, \mu = 1$ and n=50, based on 1000 samples. The dashed line is the relative efficiency using the PWM estimators and the solid line using these estimators to transform the observations and perform Pareto ML estimation.

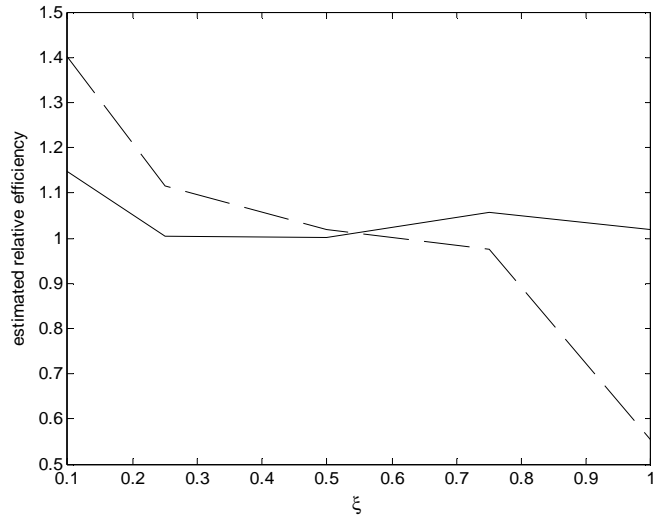

**Fig. 3** Plot of relative efficiency of estimators of $\xi$ of a GPD with $\sigma = 1, \mu = 1, n = 100$, based on 1000 samples. Estimation performed on transformed observations. The dashed line is the relative efficiency where observations were transformed using initial PWM estimates and the solid line where transformation was carried out using Zhang and Stephens (2009) estimated parameters.

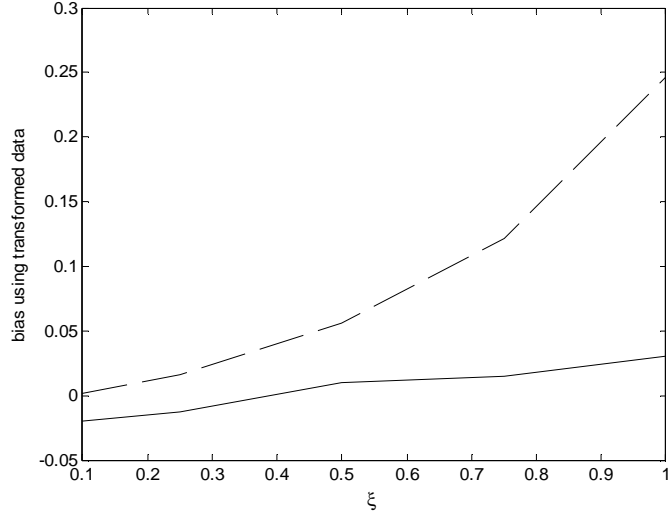

**Fig. 4** Plot of bias of estimators of $\xi$ of a GPD with $\sigma = 1, \mu = 1$, *n=100*, based on 1000 samples. Estimation performed on transformed observations. The dashed line is the bias where observations were transformed using initial PWM estimates and the solid line where the transformation was carried out using Zhang and Stephens (2009) estimated parameters.

|  | $\sigma = 2, \mu = 1$  $\xi$ | Zhang&Stephens  MSE and Bias | | Transformed (Zhang&Stephens)  MSE and Bias | | PWM  MSE and Bias | | Transformed (PWM)  MSE and Bias | |
|---|---|---|---|---|---|---|---|---|---|
| **n=50** | **0.1** | 0.0289 | -0.0220 | 0.0201 | -0.0479 | 0.0289 | 0.0334 | 0.0128 | -0.0093 |
|  | **0.25** | 0.0344 | -0.0064 | 0.0295 | -0.0174 | 0.0335 | 0.0550 | 0.0240 | 0.0362 |
|  | **0.5** | 0.0471 | 0.0128 | 0.0457 | 0.0096 | 0.0449 | 0.0989 | 0.0416 | 0.0894 |
|  | **0.75** | 0.0565 | 0.0093 | 0.0558 | 0.0078 | 0.0581 | 0.1683 | 0.0536 | 0.1593 |
|  | **1.0** | 0.0722 | 0.0268 | 0.0710 | 0.0282 | 0.1067 | 0.2894 | 0.1048 | 0.2931 |
| **n=100** | **0.1** | 0.0130 | -0.0071 | 0.0098 | -0.0187 | 0.0145 | 0.0225 | 0.0081 | 0.0020 |
|  | **0.25** | 0.0163 | -0.0041 | 0.0155 | -0.0068 | 0.0160 | 0.0283 | 0.0145 | 0.0232 |
|  | **0.5** | 0.0210 | 0.0005 | 0.0210 | -0.0005 | 0.0199 | 0.0506 | 0.0195 | 0.0460 |
|  | **0.75** | 0.0309 | 0.0120 | 0.0309 | 0.0113 | 0.0347 | 0.1257 | 0.0320 | 0.1180 |
|  | **1.0** | 0.0405 | 0.0130 | 0.0403 | 0.0133 | 0.0737 | 0.2419 | 0.0712 | 0.2437 |
| **n=250** | **0.1** | 0.0048 | -0.0053 | 0.0042 | -0.0078 | 0.0052 | 0.0068 | 0.0042 | 0.0025 |
|  | **0.25** | 0.0064 | 0.0010 | 0.0064 | 0.0003 | 0.0062 | 0.0154 | 0.0062 | 0.0142 |
|  | **0.5** | 0.0089 | -0.0027 | 0.0089 | -0.0033 | 0.0101 | 0.0247 | 0.0097 | 0.0217 |
|  | **0.75** | 0.0119 | 0.0068 | 0.0118 | 0.0063 | 0.0182 | 0.0838 | 0.0163 | 0.0774 |
|  | **1.0** | 0.0161 | 0.0027 | 0.0160 | 0.0029 | 0.0455 | 0.1911 | 0.0430 | 0.1898 |

**Table 3** Results of estimators of $\xi$ of a GPD with $\sigma = 2, \mu = 1$. MSE and estimated bias based on 1000 samples. The results of using two methods to calculate the transformed observations on which Pareto estimation is performed is shown. The procedures are the Zhang and Stephens (2009) and PWM methods.

In the following simulation study a scale parameter of 2 is used, which resulted in decreases in the MSE of the estimation using transformed observations.

| $\sigma = 2, \mu = 1$ $\xi$ | | Zhang&Stephens Relative Efficiency | Transformed (Zhang&Stephens) Relative Efficiency | PWM Relative Efficiency | Transformed (PWM) Relative Efficiency |
|---|---|---|---|---|---|
| n=50 | 0.1 | 0.8382 | 1.2032 | 0.8380 | 1.8836 |
| | 0.25 | 0.9095 | 1.0595 | 0.9327 | 1.3045 |
| | 0.5 | 0.9547 | 0.9842 | 1.0024 | 1.0821 |
| | 0.75 | 1.0850 | 1.0982 | 1.0541 | 1.1436 |
| | 1.0 | 1.1077 | 1.1260 | 0.7499 | 0.7631 |
| n=100 | 0.1 | 0.9273 | 1.2409 | 0.8348 | 1.4931 |
| | 0.25 | 0.9605 | 1.0084 | 0.9737 | 1.0798 |
| | 0.5 | 1.0689 | 1.0723 | 1.1302 | 1.1542 |
| | 0.75 | 0.9900 | 0.9921 | 0.8833 | 0.9581 |
| | 1.0 | 0.9878 | 0.9936 | 0.5428 | 0.5614 |
| n=250 | 0.1 | 1.0109 | 1.1479 | 0.9331 | 1.1510 |
| | 0.25 | 0.9774 | 0.9717 | 1.0066 | 1.0002 |
| | 0.5 | 1.0162 | 1.0149 | 0.8878 | 0.9261 |
| | 0.75 | 1.0316 | 1.0359 | 0.6742 | 0.7510 |
| | 1.0 | 0.9924 | 0.9981 | 0.3518 | 0.3723 |

**Table 4** Relative efficiency of estimators of $\xi$ of a GPD with $\sigma = 2, \mu = 1$ based on 1000 samples. The results of using two methods to calculate the transformed observations on which Pareto estimation is performed is shown.

A very special case of the three-parameter GPD is if $\sigma = \xi\mu$, where it follows that the transformation implies that x=z and the three-parameter GPD reduces to a Pareto distribution. This special case of the GPD was pointed out by (McCulloch, 1997). In the following simulation, it was checked what the result is if it is assumed that the sample is GPD distributed and it was actually purely Pareto distributed. Samples were generated from a GPD with $\sigma = \xi\mu$, and the estimation was performed assuming that it is GPD and not purely Pareto distributed.

|  | $\sigma = \mu\xi$ (Pareto) | Zhang&Stephens MSE and Bias | | Transformed (Zhang&Stephens) MSE and Bias | | PWM MSE and Bias | | Transformed (PWM) MSE and Bias | |
|---|---|---|---|---|---|---|---|---|---|
| n=50 | 0.1 | 0.0298 | -0.0106 | 0.0182 | -0.0359 | 0.0330 | 0.0449 | 0.0119 | -0.0013 |
|  | 0.25 | 0.0337 | -0.0211 | 0.0289 | -0.0219 | 0.0296 | 0.0404 | 0.0229 | 0.0324 |
|  | 0.5 | 0.0437 | 0.0056 | 0.0410 | 0.0112 | 0.0397 | 0.0958 | 0.0380 | 0.0947 |
|  | 0.75 | 0.0558 | 0.0140 | 0.0544 | 0.0228 | 0.0582 | 0.1738 | 0.0563 | 0.1719 |
|  | 1.0 | 0.0769 | 0.0406 | 0.0757 | 0.0505 | 0.1155 | 0.2988 | 0.1159 | 0.3075 |
| n=100 | 0.1 | 0.0135 | -0.0109 | 0.0097 | -0.0209 | 0.0143 | 0.0168 | 0.0080 | -0.0014 |
|  | 0.25 | 0.0157 | -0.0015 | 0.0147 | -0.0007 | 0.0153 | 0.0293 | 0.0140 | 0.0281 |
|  | 0.5 | 0.0211 | 0.0043 | 0.0208 | 0.0077 | 0.0206 | 0.0547 | 0.0203 | 0.0542 |
|  | 0.75 | 0.0318 | 0.0092 | 0.0315 | 0.0134 | 0.0341 | 0.1209 | 0.0322 | 0.1182 |
|  | 1.0 | 0.0413 | 0.0212 | 0.0409 | 0.0263 | 0.0747 | 0.2435 | 0.0754 | 0.2497 |
| n=250 | 0.1 | 0.0049 | -0.0057 | 0.0043 | -0.0078 | 0.0051 | 0.0060 | 0.0041 | 0.0025 |
|  | 0.25 | 0.0066 | -0.0045 | 0.0065 | -0.0037 | 0.0063 | 0.0084 | 0.0063 | 0.0088 |
|  | 0.5 | 0.0066 | -0.0045 | 0.0065 | -0.0037 | 0.0063 | 0.0084 | 0.0063 | 0.0088 |
|  | 0.75 | 0.0117 | 0.0045 | 0.0117 | 0.0062 | 0.0174 | 0.0838 | 0.0160 | 0.0795 |
|  | 1.0 | 0.0166 | 0.0134 | 0.0166 | 0.0154 | 0.0474 | 0.1968 | 0.0456 | 0.1966 |

**Table 5** Results of estimators of $\xi$ of a GPD with $\sigma = \xi\mu$, $\mu = 1$. MSE and estimated bias based on 1000 samples. The results of using two methods to calculate the transformed observations on which Pareto estimation is performed is shown. The procedures are the Zhang and Stephens and PWM methods.

|  | $\sigma = \mu\xi$ (Pareto) | Zhang&Stephens Relative Efficiency | Transformed (Zhang&Stephens) Relative Efficiency | PWM Relative Efficiency | Transformed (PWM) Relative Efficiency |
|---|---|---|---|---|---|
| n=50 | 0.1 | 0.8122 | 1.3315 | 0.7323 | 2.0417 |
|  | 0.25 | 0.9267 | 1.0798 | 1.0554 | 1.3647 |
|  | 0.5 | 1.0305 | 1.0984 | 1.1345 | 1.1840 |
|  | 0.75 | 1.0974 | 1.1257 | 1.0518 | 1.0876 |
|  | 1.0 | 1.0406 | 1.0563 | 0.6929 | 0.6903 |
| n=100 | 0.1 | 0.8981 | 1.2414 | 0.8434 | 1.5149 |
|  | 0.25 | 0.9973 | 1.0643 | 1.0213 | 1.1178 |
|  | 0.5 | 1.0662 | 1.0814 | 1.0936 | 1.1088 |
|  | 0.75 | 0.9617 | 0.9720 | 0.8983 | 0.9521 |
|  | 1.0 | 0.9696 | 0.9776 | 0.5354 | 0.5304 |
| n=250 | 0.1 | 0.9795 | 1.1372 | 0.9478 | 1.1718 |
|  | 0.25 | 0.9498 | 0.9571 | 0.9851 | 0.9901 |
|  | 0.5 | 0.9465 | 0.9535 | 0.8966 | 0.9285 |
|  | 0.75 | 1.0426 | 1.0470 | 0.7025 | 0.7644 |
|  | 1.0 | 0.9635 | 0.9658 | 0.3378 | 0.3509 |

**Table 6** Relative efficiency of estimators of $\xi$ of a GPD with $\sigma = \xi\mu$, $\mu = 1$ based on 1000 samples. The results of using two methods to calculate the transformed observations on which Pareto estimation is performed is shown.

In practice all the transformed value should fulfill the condition

$z = (\hat{\xi}\hat{\mu}/\hat{\sigma})x + \hat{\mu} - \hat{\xi}\hat{\mu}^2/\hat{\sigma} \geq \hat{\mu}$ and provision for this must be made when doing calculations. In the simulations performed in this work it was observed a few times in very small samples using estimated parameters to form the transformed observations, and in such a case was put equal to $\mu$. It can also be remarked that since this estimator was shown to outperform the initial estimated index iterations can possibly improve the parameters to transformation. But it was found that this leads to a very small improvement and no iterations were done in the simulation study.

## 3 Application of the transformation in POT problems

Two approaches to transform the observations above the threshold using initial estimates of the GPD parameters, such that their distribution is close to Pareto distributed is:

- Choose the threshold such that the condition $\mu = \sigma/\xi$ is fulfilled, using initial estimated parameters. This was found to be not very practical because the threshold is often so that there are too few points to perform estimation. It might work in very large samples.
- A straightforward transformation using initial estimates to transform from a GPD to Pareto Type I distribution.

Some of the estimation methods for the two parameter GPD, with $\mu = 0$, and which are applied to excesses over a threshold are probability weighted method (PWM) method, the maximum likelihood (ML) estimate and the method of Zhang and Stephens (1979). The Hill estimator of the tail index, is especially effective where the distribution is Pareto (Type I). The focus will be on POT problems for Student's t and stable distributed samples.

As was seen in (5), $\bar{F}(x) = 1 - F(x)$, where F is the distribution function of the GPD can be written as

$$\bar{F}(x) = (\mu^\alpha / x^\alpha)(\mu(1 - \mu\xi/\sigma)/x + \xi\mu/\sigma)^{-1/\xi}$$

which implies that the survival function or complementary cumulative distribution function, $\bar{F}(x) = (x/\mu)^{-1/\xi}, x \geq \mu,$ for $\mu = \sigma/\xi$. Even for $\mu \approx \sigma/\xi$ or the difference $1 - \mu\xi/\sigma$ small will lead to observations above that threshold which are approximately Pareto distributed and converges faster to Pareto distributed variables. It can be seen that as $x \to \infty$, the variables in this range are regularly varying and will the GPD distributed variables converges variables having a Pareto type I distribution too, even if $\mu \neq \sigma/\xi$. Balkema and de Haan (1974), Pickands (1975) showed that for a large group of heavy tailed distributions the excesses over a threshold are GPD distributed. This theorem is also called the Pickands–Balkema–de Haan theorem and it is also called the second theorem in extreme value theory and it is a general theorem since it can be shown that for many well know distributions the excess are Pareto distributed.

Above a suitable threshold the excesses over a threshold, variables will be distributed as a two-parameter GPD $(\mu = 0)$, of the form

$$f(x) = (1/\sigma)(1 + (\xi/\sigma)x)^{-1-1/\xi}, x \geq 0, \sigma > 0, \xi > 0.$$

Formally the theorem is (Markovich, 2007):

Theorem 1. Let $X_1, ..., X_n$ be an i.i.d. random sequence. The limit distribution of the excess distribution of the $X_i$'s is necessarily of the generalized Pareto form

$$\lim_{\mu \uparrow x_F, \mu + x < x_F} P(X_1 - \mu \mid X_1 > \mu) \to (1 + \xi x)_+^{-1/\xi}, \quad x \in R,$$

where $x_F = \sup\{x \in R : F(x) < 1\}$ is the right endpoint of the distribution F(x) and the shape parameter $\xi \in R$, $(x)_+ = x, x > 0$ and $x = 0$ otherwise.

One idea is to choose a reasonable threshold $\hat{\mu}$, estimate $\hat{\sigma}, \hat{\xi}$, and use the threshold $\hat{\mu} = \hat{\sigma}/\hat{\xi}$, and the Hill estimator can be applied on the observations larger than that $\hat{\mu}$.

Thus by choosing the threshold such that $\hat{\mu} = \hat{\sigma}\hat{\xi}$, one have observations which are approximately Pareto distributed with the same tail index. The problem with this approach is that the threshold must be less than the largest observation and also there must be enough observations between this threshold and the largest observation to find a reasonable estimate of the tail index. So there are practical issues with this approach in samples which are not very large, and the straightforward transformation approach was investigated. If this estimated threshold is larger or close to the largest observation in the sample, it might be an indication that the behavior is not yet Pareto tails, but maybe a GPD can be fitted.

The Hill estimator Hill (1975) of the index $\alpha = 1/\xi$, used for $\alpha > 0$ is defined as

$$\hat{\xi}^H(n,k) = (1/k)\sum_{j=1}^{k}\log(X_{(n-j+1)} / X_{(n-k)}),$$

where $X_{(1)} < ... < X_{(n)}$, are the ordered statistics of the sample. This implies the largest k values are used in the estimation, and the threshold is $X_{(n-k)}$. For a general threshold $\hat{\mu}$, $\hat{\xi}^H(n,k) = (1/k)\sum_{j=1}^{k}\log(X_{(n-j+1)} / \hat{\mu})$, for k observations larger than the threshold. This is the ML estimate of the inverse of the index Pareto type I distributed observations.

Alternatively, a straightforward transformation, $z/\mu = 1 + (\xi/\sigma)(x-\mu)$, can be applied on GPD distributed variables. In practice a reasonable threshold must be chosen, initial estimation performed and then ML estimation on the transformed variables which are formed using this threshold and the estimated parameters. This approach will be evaluated to estimate the tail index in POT problems. For the transformation in (3), the z's must be positive and such a check should also be done, when transforming and using estimated parameters.

Samples were generated from, a threshold chosen, and the excesses over the threshold, the transformed observations $z_{(j)} = x_{(j)} - \hat{\mu}$, $j = 1,...,n$, were formed to perform the initial estimation to estimate the threshold such that the observations are closest to Pareto distributed. The threshold were chosen so that there are k=100 observations larger than the threshold.

Using these excesses as the sample, the following four estimators were compared: The Hill estimator for the chosen threshold, The Zhang and Stephens (2009 and the ML estimation of the shape parameter using excesses, the Hill estimator, Pareto ML estimation on transformed observations using the estimated parameters of the Zhang and Stephens (2009) method to perform the transformation to Pareto distributed variables.

The number of simulations in each case is m=1000, and MSE was calculated as $(1/m)\sum_{j=1}^{m}(\hat{\xi}_j - \xi)^2$ for the various estimators. The bias is calculated as the true parameter minus the average estimated parameter.

Three thresholds were tested, using the largest, k=50,100,250 for sample sizes of 1000, 2500 and 5000.observations and different values for the tail index of the simulated samples. The data simulated were assumed to be from symmetric distributions and POT threshold estimation for the stable and Student's t-distributions will be considered. For the t-distribution the tail index is the degrees of freedom (McNeill and Saladin, 1997) give details how to derive this result.

|  | Student's t df | Zhang&Stephens Bias and Mse | | ML Bias and Mse | | Hill Bias and Mse | | Transformed (Z&S) Bias and Mse | |
|---|---|---|---|---|---|---|---|---|---|
| n=1000 k=100 | 1 | 0.0259 | 0.0357 | 0.0286 | 0.0384 | -0.0027 | 0.0095 | 0.0259 | 0.0357 |
|  | 2 | 0.0267 | 0.0239 | 0.0485 | 0.0275 | -0.0419 | 0.0047 | 0.0267 | 0.0239 |
|  | 3 | 0.0373 | 0.0180 | 0.0657 | 0.0223 | -0.0735 | 0.0069 | 0.0369 | 0.0177 |
|  | 4 | 0.0570 | 0.0188 | 0.0901 | 0.0251 | -0.0990 | 0.0108 | 0.0532 | 0.0165 |
|  | 5 | 0.0586 | 0.0186 | 0.0938 | 0.0253 | -0.1174 | 0.0146 | 0.0507 | 0.0146 |
| n=2500 k=100 | 1 | 0.0191 | 0.0407 | 0.0219 | 0.0438 | -0.0023 | 0.0110 | 0.0191 | 0.0407 |
|  | 2 | 0.0112 | 0.0212 | 0.0325 | 0.0239 | -0.0158 | 0.0027 | 0.0112 | 0.0212 |
|  | 3 | 0.0286 | 0.0178 | 0.0573 | 0.0218 | -0.0362 | 0.0026 | 0.0285 | 0.0177 |
|  | 4 | 0.0331 | 0.0154 | 0.0653 | 0.0199 | -0.0541 | 0.0037 | 0.0309 | 0.0141 |
|  | 5 | 0.0401 | 0.0151 | 0.0748 | 0.0204 | -0.0683 | 0.0053 | 0.0357 | 0.0130 |
| n=5000 k=100 | 1 | 0.0187 | 0.0413 | 0.0214 | 0.0446 | 0.0012 | 0.0092 | 0.0187 | 0.0413 |
|  | 2 | 0.0065 | 0.0219 | 0.0271 | 0.0245 | -0.0078 | 0.0029 | 0.0065 | 0.0219 |
|  | 3 | 0.0123 | 0.0183 | 0.0403 | 0.0213 | -0.0215 | 0.0017 | 0.0119 | 0.0180 |
|  | 4 | 0.0243 | 0.0160 | 0.0563 | 0.0200 | -0.0363 | 0.0020 | 0.0227 | 0.0151 |
|  | 5 | 0.0240 | 0.0148 | 0.0581 | 0.0189 | -0.0481 | 0.0028 | 0.0196 | 0.0125 |

**Table 7** MSE and bias of estimators of $\xi$ of a GPD based on the largest 100 observations in each sample. There were 1000 samples generated from a standard Student's t distribution of sample n each.

|  | Stable Index | Zhang&Stephens Bias and Mse | | ML Bias and Mse | | Hill Bias and Mse | | Transformed (Z&S) Bias and Mse | |
|---|---|---|---|---|---|---|---|---|---|
| n=1000 k=100 | 1.9 | 0.2144 | 0.0694 | 0.2381 | 0.0822 | 0.2427 | 0.0600 | 0.2127 | 0.0674 |
|  | 1.7 | 0.0190 | 0.0240 | 0.0348 | 0.0266 | 0.1593 | 0.0278 | 0.0190 | 0.0240 |
|  | 1.5 | 0.0055 | 0.0265 | 0.0192 | 0.0289 | 0.0866 | 0.0115 | 0.0055 | 0.0265 |
|  | 1.3 | 0.0082 | 0.0307 | 0.0188 | 0.0332 | 0.0344 | 0.0067 | 0.0082 | 0.0307 |
|  | 1 | 0.0277 | 0.0402 | 0.0307 | 0.0432 | -0.0035 | 0.0098 | 0.0277 | 0.0402 |
| n=2500 k=100 | 1.9 | 0.0165 | 0.0238 | 0.0330 | 0.0264 | 0.2478 | 0.0627 | 0.0165 | 0.0238 |
|  | 1.7 | -0.0377 | 0.0267 | -0.0217 | 0.0278 | 0.1070 | 0.0142 | -0.0377 | 0.0267 |
|  | 1.5 | -0.0098 | 0.0286 | 0.0044 | 0.0306 | 0.0453 | 0.0063 | -0.0098 | 0.0286 |
|  | 1.3 | 0.0008 | 0.0294 | 0.0111 | 0.0317 | 0.0174 | 0.0059 | 0.0008 | 0.0294 |
|  | 1 | 0.0153 | 0.0370 | 0.0174 | 0.0398 | -0.0023 | 0.0090 | 0.0153 | 0.0370 |
| n=5000 k=100 | 1.9 | -0.1091 | 0.0364 | -0.0955 | 0.0355 | 0.2099 | 0.0457 | -0.1091 | 0.0364 |
|  | 1.7 | -0.0279 | 0.0275 | -0.0114 | 0.0290 | 0.0603 | 0.0067 | -0.0279 | 0.0275 |
|  | 1.5 | -0.0038 | 0.0283 | 0.0105 | 0.0307 | 0.0210 | 0.0046 | -0.0038 | 0.0283 |
|  | 1.3 | 0.0136 | 0.0293 | 0.0244 | 0.0320 | 0.0139 | 0.0063 | 0.0136 | 0.0293 |
|  | 1 | 0.0160 | 0.0375 | 0.0186 | 0.0402 | 0.0047 | 0.0096 | 0.0160 | 0.0375 |

**Table 8** MSE and bias of estimators of $\xi$ of a GPD based on the largest 100 observations in each sample. There were 1000 samples generated from a symmetric stable distribution of sample n each.

It can be seen there is a slight improvement in the estimation results for lighter tails, and that the Hill's estimator performs best when the tail are very heavy, and close to one. But this may also depend on the scale and shape parameter. In many cases the Hill estimator has a small MSE with a large bias.

## 4 Conclusions and remarks

It was found that the transformation method leads to estimation results which are at the worst just as good as the original method to perform the transformation, and in many cases to improved estimation results.

Since there are so many methods of estimation, it might take much research to find the optimal way to transform the transformation.